\documentclass[preprint,superscriptaddress,11pt]{revtex4-1}
\usepackage{graphicx}
\usepackage{dcolumn}
\usepackage{color}
\usepackage{times}
\usepackage{bm}
\usepackage{amssymb}
\usepackage{amsmath}
\usepackage{epsfig}
\usepackage{epstopdf}
\usepackage{dsfont}

\begin{document}
\renewcommand{\thefootnote}{\fnsymbol {footnote}}

\title{Effects of Hawking radiation on the entropic uncertainty in a Schwarzschild space-time}

\author{Dong Wang} \email{dwang@ahu.edu.cn (D.W.)}
\affiliation{School of Physics \& Material Science, Anhui University, Hefei
230601, China}
\affiliation{CAS Key Laboratory of
Quantum Information, University of Science and
Technology of China, Hefei 230026, China}

\author{Wei-Nan Shi}
\affiliation{School of Physics \& Material Science, Anhui University, Hefei
230601, China}

\author{Ross D. Hoehn}
\affiliation{Department of Chemistry, Department of Physics
and Birck Nanotechnology Center, Purdue University, West Lafayette, IN 47907 USA}

\author{Fei Ming}
\affiliation{School of Physics \& Material Science, Anhui University, Hefei
230601, China}

\author{Wen-Yang Sun}
\affiliation{School of Physics \& Material Science, Anhui University, Hefei
230601, China}

\author{Sabre Kais}
\affiliation{Department of Chemistry, Department of Physics
and Birck Nanotechnology Center, Purdue University, West Lafayette, IN 47907 USA}
\affiliation{Santa Fe Institute, Santa Fe, NM 87501 USA}

\author{Liu Ye}\email{yeliu@ahu.edu.cn (L.Y.)}
\affiliation{School of Physics \& Material Science, Anhui University, Hefei
230601, China}

\date{\today}

\begin{abstract}{\bf
Heisenberg uncertainty principle describes a basic restriction on observer¡¯s ability of precisely predicting the
measurement for a pair of non-commuting observables, and virtually is at the core of quantum mechanics. We
herein aim to study entropic uncertainty relation under the background of the Schwarzschild black hole and
its control. Explicitly, we develop dynamical features of the measuring uncertainty via entropy in a practical
model where a stationary particle interacts with its surrounding environment while another particle --- serving
as a quantum memory reservoir --- undergoes freefall in the vicinity of the event horizon of the Schwarzschild
space-time. It shows higher Hawking temperatures would give rise to an inflation of the entropic uncertainty
on the measured particle. This is suggestive the measurement uncertainty is strongly correlated with degree of mixing present in the evolving particles. Additionally, based on information flow theory, we provide a physical interpretation for the observed dynamical behaviors related with the entropic uncertainty in such a realistic scenario. Finally, an efficient strategy is proposed to reduce the uncertainty by non-tracing-preserved operations. Therefore, our explorations may improve the understanding of the dynamic entropic uncertainty in a
curved space-time, and illustrate predictions of quantum measurements in relativistic quantum information
sciences.}
\end{abstract}

\maketitle

\section{Introduction}

The Heisenberg uncertainty principle \cite{Heisenberg} --- being related to precision of simultaneous measurements for a pair of non-commutative observables ---
is a foundational feature and fundamental insight of quantum theory \cite{PatrickJ}. A later generalization was developed by Kennard \cite{E.H. Kennard} and Robertson \cite{H. P. Robertson}
into a standard deviation $\Delta{ {\cal S}}\Delta{{\cal R}}\geq |\langle[{{\cal S}},{\cal R}]\rangle|/2$ for a pair of arbitrary incompatible observables ${\cal S}$ and ${\cal R}$ \cite{ Lorenzo, Kunkun,Kunkun2,qiao}.
Notably, the standard deviation is not optimal for the quantification of the uncertainty because the lower bound is state-dependent.
At present, the more popular method to depict the uncertainty principle is to employ entropic measures other than that mentioned deviation. Denoting the outcome $\epsilon$ with probability $p(\epsilon)$ for a measurement $\varepsilon$,
$H(\varepsilon)=-\sum_\epsilon p(\epsilon) {\rm log}_2 p(\epsilon)$ is denoted as the Shannon entropy, which features the uncertainty
regarding $\varepsilon$ before we are aware of the outcome for the measurement. With respect to ${\cal R}$ and ${\cal R}$, the entropic
uncertainty can be mapped into the inequality $H({{\cal S}})+H({\cal R})\geq {\rm log}_2{\frac1c}$ \cite{D. Deutsch, K. Kraus,H. Maassen},
$c=\underset{\epsilon,\tau}{\rm max}\{|\langle\Psi_\epsilon| \Phi_\tau\rangle|^2\}$ is the maximal overlap, where $|\Psi_\epsilon\rangle$ and $|\Phi_\tau\rangle$ stand for the corresponding eigenvectors of ${\cal S}$ and ${\cal R}$.
Owing to $c$ being independent of system state, the entropic uncertainty relation (EUR) enables us to better quantify the measured uncertainty compared with the standard deviation.

Nevertheless, the uncertainty relation can be circumvented if the measured object is entangled with another one (serving as a so-called quantum memory reservoir).
Notably, an observer can perfectly predict the measurement outcome if his/her particle
is maximally entangled with the observed particle. This can be interpreted through a
quantum-memory-assisted entropic uncertainty relation (QMA-EUR), which was firstly proposed by \cite{Joseph} and then supported by \cite{M. Berta}. This relation can be expressed as:
\begin{align}
H({\cal S}|B)+H({\cal R}|B)\geq H(A|B)+ {\rm log}_2 {\frac1c},
\label{Eq.3}
\end{align}
$H({\cal S}|B)=S(\hat{\rho}_{{\cal S}B})-S(\hat{\rho}_B)$ is denoted as the conditional von Neumann
entropy \cite{C. F. Li,R. Prevedel}, where $\hat{\rho}_{{\cal S}B}=\sum\limits_\epsilon\left (|\Psi_\epsilon\rangle_A\langle\Psi_\epsilon|\otimes{\mathds{1}}_B\right)\hat{\rho}_{AB} \left(|\Psi_\epsilon\rangle_A\langle\Psi_\epsilon
|\otimes{\mathds{1}}_B\right )$. It is noteworthy that the most promising form of Heisenberg's relation might also
be interpreted as a so-called uncertainty game between two participants (Alice and Bob). In the beginning, Bob produces particle $A$,
correlated with particle $B$, which acts as a quantum memory. Bob then sends
$A$ to Alice at a distant site, she then implements one of two possible measurements and broadcasts her determination to Bob.
This action will enable Bob to predict Alice's measured results within minimal uncertainty.
Remarkably, those measured outcomes can be precisely guessed when particle $A$ is maximally entangled with $B$.
On the other hand, if $A$ and $B$ are unentangled, the QMA-EUR will naturally reduce to the EUR \cite{D. Deutsch} because of $H({\cal S})\geq H({\cal S}|B)$ and $H({\cal R})\geq H({\cal R}|B)$. Besides,
a tighter bound ${\rm log}_2 {\frac1c}+H(A)$ is obtained without quantum memory compared to EUR's bound.
To date, serval authors have investigated the characteristics of QMA-EUR in terms of R\'{e}nyi entropy \cite{P.J. Coles, M. Tomamichel,J.zhang1}, and there exist several studies employing this relation as their measure of uncertainty \cite{P. J. Coles,Pati,Xiao,F. Adabi,J. Zhang,M. Hu,M. Hu2,S. Liu,J. Schneeloch,M. L. Hu,F. Adabi2,A.E. Rastegin,M. Berta2,R. Adamczak,K. Baek}.

In practice, a physical system is remarkably susceptible to ambient environments, which may unavoidably inject decoherence effects.
In this regard, it is of fundamental importance to pursue an understanding of the influence of various decoherence effects on QMA-EUR;
these effects have been shown to be nontrivial in the domain of quantum information science \cite{Z. Y. Xu,H. M. Zou,Ai-Jun Huang,Dong2,Dong1,Yu Min}.
Recently, Fan {\it {et al.}} \cite{Jun Feng1,Jun Feng2}, and Jing {\it et al.} \cite{Lijuan Jia} have explored the dynamic characteristics of the QMA-EUR
while the particle to be measured is taken to be curved space-time (\emph{e.g.} Schwarzschild and de Sitter spaces); Huang {\it et al.} \cite{huang}
demonstrated a tighter bound of the entropic uncertainty relation based on the Holevo quantity in Schwarzschild black hole.
However, previous investigations mainly
focused on ideal models where the measured particle of interest is free from its surrounding environments when the other one --- taking on the role of quantum memory reservoir ---
is under either a flat or a curved space-time. Therefore, it is of great interest to clarify how both a noisy environment and the inclusion of Hawking radiation effects act on the measurement precision in entangled systems.

In this article, our purpose is to examine the dynamic behaviors of the QMA-EUR in the presence of and under the influence of the additive effects of quantum noises and Hawking radiation, respectfully.
It has been found that the lower bound of the uncertainty relies --- not only on the measurement performed by the observer, but --- on both environment
and the Hawking temperature ($T$) from the black hole are responsible for deforming space-time local to the particles. Firstly, we take into account a system consisting of bipartite $AB$ where Particle $A$
stays at an open system featured by nonunital environment noises, while Particle $B$ is hovering on the event horizon of the Schwarzschild black hole
relative with $A$; we then determine the associated measurement uncertainties on the particle $A$. Secondly, another bipartite correlated system is considered
where $A$ is suffering from unital noises and $B$ is in the near event horizon of a Schwarzschild black hole. Additionally, the dynamic features of the
measurement uncertainty regarding a pair of non-commutative observables are analysed. Finally, we suggest a simple functional method for steering the measurement uncertainty
by using a series of local non-tracing-preserved operations.

\section{Hawking effect in a Schwarzschild space-time}

Let us recall the definition for metric Hawking radiation within the Dirac field model in Schwarzschild space-time. Conventionally, the Dirac equation with respect to general curved space-time
is described through $[\gamma^ae_a^\mu(\partial_\mu+\Gamma_\mu)]\Psi=0$ \cite{D. R. Brill,Jing1},
with the Dirac matrix $\gamma^a$, the mass of the Dirac field $\mu$, $e^{\mu}_{a}$ the inverse of the tetrad $e_{\mu}^a$, and spin connection given by
$\Gamma_{\mu}=\frac18[\gamma^a,\gamma^b]e_a^\nu e_{b\nu;\mu}$. In the Schwarzschild background, the metric can be specified as:
\begin{align}
ds^2=-(1-\frac{2M}r)dt^2+(1-\frac{2M}r)^{-1}dr^2+r^2({\rm sin}^2\theta d\varphi^2+d\theta^2),
\end{align}
with the mass of the black hole given by $M$.
Then, we can obtain the Dirac equation within the context of Schwarzschild space-time:
\begin{align}
-\frac{\gamma_0}{\sqrt{1-\frac{2M}r}}\frac{\partial\Psi}{\partial t}+\gamma_1\sqrt{1-\frac{2M}r}[\frac{\partial}{\partial t}+\frac1r+\frac M{2r(r-2M)}]\Psi+\frac{\gamma_2}{r}\frac{\partial\Psi}{\partial\theta}+\frac{\gamma_3}{r \rm{sin}\theta}\frac{\partial\Psi}{\partial\varphi}=0.
\label{Eq.7}
\end{align}
Based on the above formula, one could derive the positive (fermions) frequency outgoing solutions as:
\begin{align}
&\Psi_k^{{\rm I}+}=\aleph e^{-i\varpi u},\ (r<r_+), \nonumber \\
&\Psi_k^{{\rm II}+}=\aleph e^{i\varpi u},\ (r>r_+),
\label{Eq.8}
\end{align}
for both the outside and inside regions for the event horizon, with a 4-component Dirac spinor $\aleph$ given by:
\begin{align}\aleph=\left(
\begin{array}{c}
if(r)\phi_{nm}^{\pm}(\theta,\varphi) \\
f(r)\phi_{nm}^{\pm}(\theta,\varphi)
\end{array}
\right) ,
\end{align}
where
\begin{align}
f(r)&=(r^4-2Mr^3)^{-1/4}, \nonumber \\
\phi_{nm}^{+}&=\left( \begin{array}{c}
\sqrt{\frac{n+m}{2n}}Y_l^{m-\frac12}\\
\sqrt{\frac{n-m}{2n}}Y_l^{m+\frac12}
\end{array}
\right)\ {\rm for}\ n=l+\frac12, \nonumber \\
\phi_{nm}^{-}&=\left(
\begin{array}{c}
\sqrt{\frac{n-m+1}{2n}}Y_l^{m-\frac12}\\
-\sqrt{\frac{n+m+1}{2n}}Y_l^{m+\frac12}
\end{array}
\right) {\rm for}\ n=l-\frac12.
\end{align}
Within the above, $\varpi$ denotes the homochromous frequency of the Dirac flied, $u=t-\overline{r}$ with $\overline{r}= {\rm ln}[\frac{r-2M}{2M}+r]^{2M}$
Representing the tortoise coordinate. Particles and antiparticles would be sorted by means of the future-directed time-like Killing vector within corresponding regions.

To further probe the evolution of a Dirac particle in such scenario, the light-like Kruskal coordinates addressing Schwarzschild space-time may be imposed as:
\begin{align}
u=-4M\ {\rm ln}(\frac U{4M}), v=4M\ {\rm ln}(\frac V {4M}),\ {\rm for}\ r<r_+, \nonumber \\
u=-4M\ {\rm ln}(\frac {-U}{4M}), v=4M\ {\rm ln}(\frac V {4M}),\ {\rm for}\ r>r_+.
\end{align}
Then one can construct a set of complete basis vectors with respect to all the positive-energy modes by performing an analytic
continuation to equation (\ref{Eq.8}) in accordance with the claims made in Ref. \cite{T. Damoar}. We can look for
an alternative set of complete basis through performing an analytic continuation of equation (\ref{Eq.7}) as:
\begin{align}
&\vartheta_k^{{\rm I}+}=e^{2\pi\omega_kM}\Psi_k^{{\rm I}+}+e^{-2\pi\omega_kM}\Psi_{-k}^{\rm II-}, \nonumber \\
&\vartheta_k^{{\rm II}+}=e^{2\pi\omega_kM}\Psi_k^{\rm II+}+e^{-2\pi\omega_kM}\Psi_{-k}^{\rm I-}. 
\end{align}
Afterwards, the Bogoliubov transformations \cite{S.M. Barnett} can be performed to transform between creation and annihilation operators in Schwarzschild and Kruskal coordinates.
As a consequence, one can express the ground state and excited states for the Kruskal particle with the mode $k$ are given by:
\begin{align}
&|0\rangle_k^+\rightarrow a|0_k\rangle_{\rm I}^+|0_{-k}\rangle_{\rm II}^-+b |1_k\rangle_{\rm I}^+|1_{-k}\rangle_{\rm II}^-, \nonumber \\
&|1\rangle_k^+\rightarrow |1_k\rangle_{\rm I}^+|0_{-k}\rangle_{\rm II}^- ,
\label{Eq.191}
\end{align}
on the basis of Schwarzschild space-time, where:
\begin{align}
a=\frac1{\sqrt{1+e^{-\omega_k/T}}}, \ \ \ \ \ \ b=\frac1{\sqrt{1+e^{\omega_k/T}}} .
\label{Eq.10}
\end{align}
$T=\frac1{8\pi M}$ is the Hawking temperature \cite{J.C. Wang}, $|\alpha_k\rangle_{\rm I}^+$ and $|\alpha_{-k}\rangle_{\rm II}^-$ are denoted as the orthonormal basis for the outside and inside regions of the event horizon, respectively. For simplicity, we impose that $\omega_k=\omega$, $|\alpha_k\rangle_{\rm I}^+$ ($|\alpha_{-k}\rangle_{\rm II}^-$) will be denoted as $|\alpha\rangle_{\rm I}$ ($|\alpha\rangle_{\rm II}$) hereafter.

\section {QMA-EUR in the curved space-time}

As stated during the uncertainty game, Bob first transmits particle $A$ to Alice; this particle is initially correlated with his particle $B$, and acts as a quantum memory. Alice performs her measurement of either ${\cal R}$ or ${\cal S}$ on $A$ and informs Bob of the
choice of measurement through classical messages; upon receiving the classical information, Bob is capable of performing a minimization over the uncertainty concerning
the measurement result of $A$. To explore the congregate impacts of Hawking effect and quantum noises on the entropic uncertainty, we herein
take into account a two-particle system which initially shares generic Bell-diagonal states:
\begin{align}
\hat{\rho}_{AB}=\frac14({{\mathds{1}}}_A\otimes {{\mathds{1}}}_B+\sum_{i=1}^3c_i\hat{\sigma}_i^A\otimes\hat{\sigma}_i^B),
\label{eq.13}
\end{align}
which is characterized by including both mixed and pure states within the system's Hilbert space. The correlation coefficient $c_i={\rm Tr}_{AB}(\hat{\rho}_{AB}\hat{\sigma}_i^A\otimes\hat{\sigma}_i^B)$ satisfies $0\leq |c_i|\leq1$, and ${{{\mathds{1}}}}$ and ${\hat{\sigma}_i}$ being an identity operator and a Pauli matrix, respectively. Incidentally, the state naturally reduces to maximal entanglement under the condition $|c_i|=1$. Suppose the state of the particle held by Alice is built on the field modes $r$, only perceived by her own detector; yet the state of Bob's particles would be abridged from mode $s$,
to which only his detector is sensitive. As a result, the implementation of
$A$ and $B$ as superscripts would exhibit a double meaning, both as particle host/observer and as the constituent field modes for the particle state. After their initial entanglement, we assume $B$ is housed at a
Schwarzschild space-time located near the event horizon while Alice's Particle, $A$, is maintained in a static, flat space-time.

Quantum noises can typically be classified into two categories: unital and nonunital.
Conventionally, a unital noise satisfies the unital condition $\Upsilon_w^A(\frac1d\mathds{1}_A)=\frac1d\mathds{1}_A$
with $\Upsilon_w^A(\hat{\rho}_A)=\sum_w\hat{{K}}_w\hat{\rho}_A\hat{{K}}_w^{\dag}$; where the Kraus operators $\hat{{K}}_w$ is employed to describe
noise in mathematical way. Actually, there exists a few canonical categories of unital noise: bit-flip (BF), bit-phase-flip (BPF), and phase damping (PD). We define the noise as nonunital if this condition is not met, this includes:
amplitude damping, depolarizing (DP), etc.

\subsection{The dynamics of QMA-EUR under collective effects of Hawking radiation field and nonunital noises}

In a realistic setting, a quantum system is essentially open and inevitably suffers from interactions with any surrounding quantum noise, resulting in the decoherence of the system.
Taking into account a scenario where qubit $A$ stays at an inertial system and experiences a canonical and nonsemiclassical class of quantum noise (depolarizing), and
Particle $B$ --- acting as quantum memory --- resides in a Dirac field in the presence of Hawking temperature $T$.

Typically, a nonunital noise can be modeled by a depolarizing noise in a thermal field, whose Kraus operators can be depicted as:
\begin{align}
\hat{ K}^0=
\sqrt {1-p}\left(
\begin{array}{cc}
1 & 0 \\
0 & 1
\end{array}
\right),\
\hat{ K}^1=
\sqrt {\frac p3}\left(
\begin{array}{cc}
0 & 1\\
1 & 0
\end{array}
\right),\nonumber \\
\hat{ K}^2=
\sqrt {\frac p3}\left(
\begin{array}{cc}
0 & -i \\
i & 0
\end{array}
\right),\
\hat{ K}^3=\sqrt {\frac p3}\left(
\begin{array}{cc}
1 & 0 \\
0 & -1
\end{array}
\right),
\end{align}
where $p=1-e^{-\delta t}$ $(\delta=(1+\frac2{e^{-h\omega/k_BT}-1})\delta_0)$ is the noisy factor. $\delta$ denotes the energy relaxation rate,
$hw$ represents the systematic transition energy, with the storage time $t$ and the Boltzman constant $k_B$.
As a result, the systematic state of the bipartite system when exposed to noise is written as:
\begin{align}
\hat{\rho}_{AB_{\rm I}B_{\rm II}}=\overset{3}{\underset{u=0}\sum} ({\hat{K}^u_A\otimes {\mathds{1}}_{B_{\rm I}}\otimes {\mathds{1}}_{B_{{\rm II}}})\hat{\rho}_{AB_{\rm I}B_{\rm II}}(0) (\hat{K}^u_A \otimes {{{\mathds{1}}}_{B_{\rm I}}}\otimes {{\mathds{1}}}_{B_{\rm II}})^{\dag}},
\label{Eq.16}
\end{align}
with the systematic state being:
\begin{align}
\hat{\rho}_{AB_{\rm I}B_{\rm II}}(0)&=\frac{1+c_3}{4}(a^2|000\rangle\langle000|+ab|000\rangle\langle011|+ab|011\rangle\langle000|+b^2|011\rangle\langle011|+|110\rangle\langle110|) \nonumber \\
&+\frac{1-c_3}{4}(a^2|100\rangle\langle100|+ab|100\rangle\langle111|+ab|111\rangle\langle100|+b^2|111\rangle\langle111|+|010\rangle\langle010|) \nonumber \\
&+\frac{1-c_2}{4}(a|000\rangle\langle110|+b|011\rangle\langle110|+a|110\rangle\langle000|+b|110\rangle\langle011|) \nonumber \\
&+\frac{c_1+c_2}{4}(a|010\rangle\langle100|+b|010\rangle\langle111|+a|100\rangle\langle010|+b|111\rangle\langle010|), \nonumber
\end{align}
after interaction with the radiation effect.

The state of qubit $B$ --- shown in Eq. (\ref{Eq.16}) --- could be transformed via Eq. (\ref{Eq.191}), and after all degrees of freedom in physical inaccessible Region II are traced over,
the bipartite state reduces to:
\begin{align}
{\hat{\rho}}_{AB_\mathrm{I}}={\rm Tr}_{B_{\rm II}}({\hat{\rho}}_{AB_\mathrm{I}B_{\rm II}})=\left(
\begin{array}{cccc}
{\hat{\rho}}_{11} & {\hat{\rho}}_{12} & {\hat{\rho}}_{13} & {\hat{\rho}}_{14} \\
{\hat{\rho}}_{21} & {\hat{\rho}}_{22} & {\hat{\rho}}_{23} & {\hat{\rho}}_{24} \\
{\hat{\rho}}_{31} & {\hat{\rho}}_{32} & {\hat{\rho}}_{33} & {\hat{\rho}}_{34} \\
{\hat{\rho}}_{41} & {\hat{\rho}}_{42} & {\hat{\rho}}_{43} & {\hat{\rho}}_{44}\\
\end{array}
\right);
\label{eq.19}
\end{align}
where,
\begin{align}
&{\hat{\rho}}_{11}=\frac1{12} (2 p - a^2 (1 + c_3) (-3 + 2 p)),\nonumber \\
&{\hat{\rho}}_{22}=\frac1{12} (6 - 2 p + a^2 (1 + c_3) (-3 + 2 p)),\nonumber \\
&{\hat{\rho}}_{33}=\frac1{12} (2 p + a^2 (-1 + c_3) (-3 + 2 p)),\nonumber \\
&{\hat{\rho}}_{44}=\frac1{12} (6 - 2 p - a^2 (-1 + c_3) (-3 + 2 p)),\nonumber \\
&{\hat{\rho}}_{14}={\hat{\rho}}_{41}=\frac1{12} a (c_1 (3 - 2 p) + c_2 (-3 + 4 p)),\nonumber \\
&{\hat{\rho}}_{23}={\hat{\rho}}_{32}=\frac1{4} a (c_1 + c_2) - \frac1{6} a (c_1 + 2 c_2) p,\nonumber \\
&{\hat{\rho}}_{12}={\hat{\rho}}_{13}={\hat{\rho}}_{21}={\hat{\rho}}_{31}={\hat{\rho}}_{42}={\hat{\rho}}_{43}={\hat{\rho}}_{24}={\hat{\rho}}_{34}=0.
\label{eq.20}
\end{align}

To expose the dynamic features of the measured uncertainty under such a scenario, we adopt ($\hat{\sigma}_x,\hat{\sigma}_z$) as a pair of incompatible measurements.
Consequently, the entropies given in Eq. (\ref{Eq.3}) could yield:
\begin{align}
{\hat{\rho}}_{\hat{\sigma}_xB_\mathrm{I}}=&\frac {a^2 (3 - 2 p) + 2 p}{12}(|00\rangle\langle00|+|10\rangle\langle10|)+\frac{ 6 - 2 p + a^2 (-3 + 2 p)}{12}(|01\rangle\langle01|+|11\rangle\langle11|)\nonumber \\
&+\frac{a c_1 (3 - 2 p)}{12}(|01\rangle\langle10|+|10\rangle\langle01|+|00\rangle\langle11|+|11\rangle\langle00|),\nonumber\\
{\hat{\rho}}_{\hat{\sigma}_zB_\mathrm{I}}=&\frac{2 p - a^2 (1 + c_3) (-3 + 2 p)}{12}(|00\rangle\langle00|+|10\rangle\langle10|)\nonumber\\
&+\frac{ 6 - 2 p + a^2 (1 + c_3) (-3 + 2 p)}{12}(|01\rangle\langle01| +|11\rangle\langle11|).
\label{Eq.21}
\end{align}
In terms of the eigenvalues of post-measured states, the von Neumann entropies can be analytically given by:
\begin{align}
H({\hat{\rho}}_{\hat{\sigma}_xB_\mathrm{I}})=H_{\rm bin}(\frac{3+\lambda}6)+1, \quad
H({\hat{\rho}}_{\hat{\sigma}_zB_\mathrm{I}})=-\sum_{i}\lambda_{i} {\rm log}_2\lambda_{i} .
\end{align}
Where within the above, $H_{\rm bin}$ denotes the binary entropy, where $H_{\rm bin}(x)=-x{\rm log}_2x-(1-x){\rm log}_2(1-x)$,
$\lambda=\{[1 + a^4 +a^2 (-2 + {c_1}^2)](3 - 2 p)^2\}^{1/2}$,
$\lambda_1=[2 p - a^2 (1 + c_3) (-3 + 2 p)]/12 $,
$\lambda_2=[6 - 2 p + a^2 (1 + c_3) (-3 + 2 p)]/12 $,
$\lambda_3=[2 p + a^2 (-1 + c_3) (-3 + 2 p)]/12 $ , and
$\lambda_4=[6 - 2 p - a^2 (-1 + c_3) (-3 + 2 p)]/12$.

\begin{figure*}
\centering
\includegraphics[width=13cm]{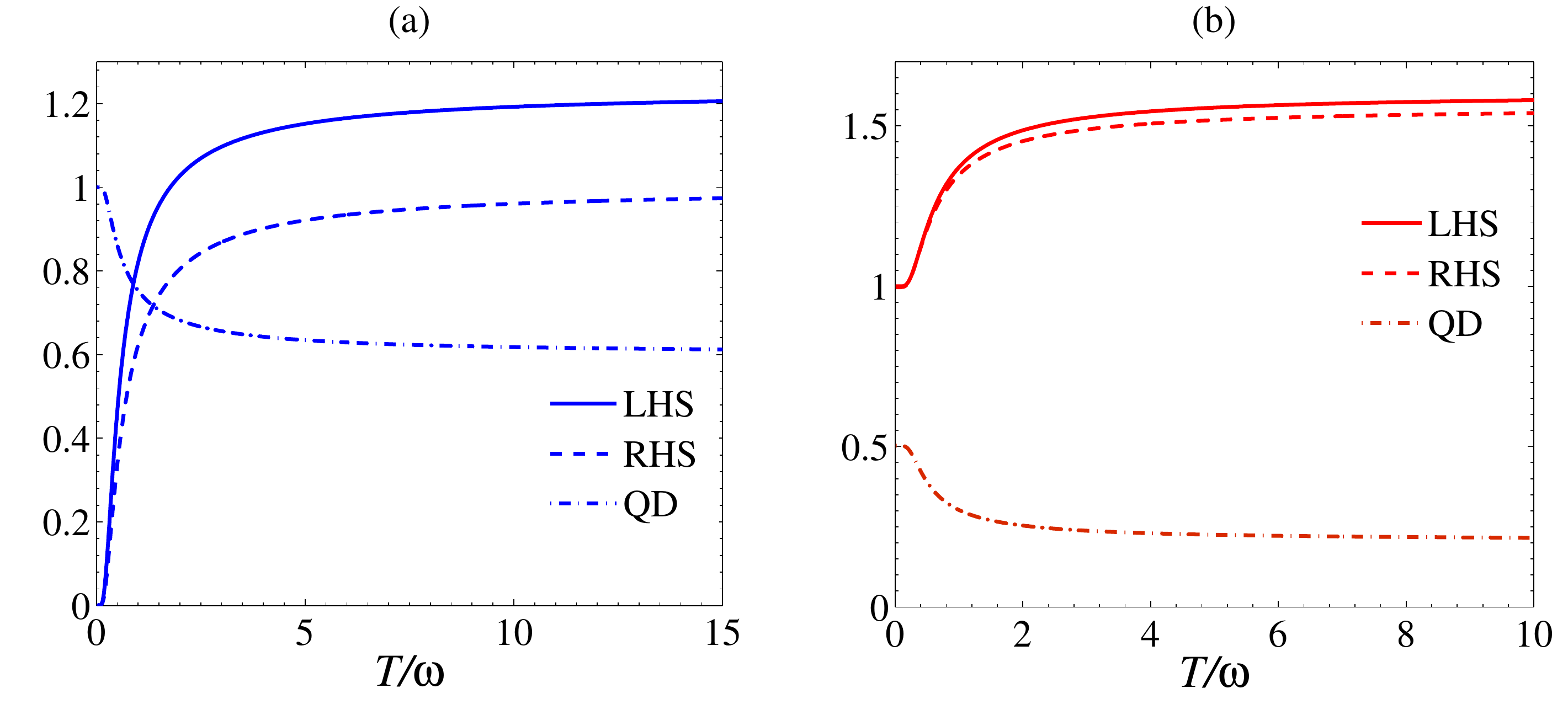}
\caption{Uncertainty and quantum discord plotted versus $T/\omega$ when particle $A$ is experiencing the effects of a depolarizing
noise and $B$ is located near the event horizon in a Schwarzschild space-time. LHS and RHS are in reference to. Eq. (\ref{Eq.3}). QD represents quantum discord of $AB$, tracing over Region $\rm II$ of $B$.
(a): $p=0$ corresponding to Particle $A$ being
isolated from any noise and $(c_1,c_2,c_3)=(1,-1,1)$; (b): $p=0.2$ and $(c_1,c_2,c_3)=(0.9,0.8,-0.9)$.}
\label{fig.1}
\end{figure*}

We obtain $H({\hat{\rho}}_{B_\mathrm{I}})=H_{\rm bin}(\frac{ a^2 (3 - 2 p) + 2 p}6)$, by calculating ${\hat{\rho}}_{B_\mathrm{I}}={\rm Tr}_A({\hat{\rho}}_{AB_\mathrm{I}})$.
Thus, we easily derive that the measurement uncertainty quantified by Eq. (\ref{Eq.3}) explicitly is:
\begin{align}
U=H_{\rm bin}(\frac{3+\lambda}6)-\sum_{s=1}^4\lambda_{i} {\rm log}_2\lambda_{i}-2H_{\rm bin}(\frac{ a^2 (3 - 2 p) + 2 p}6)+1.
\end{align}
From the uncertainty's analytic expression, one can realize the uncertainty of interest is not only related to the noisy factor $p$, but also
$a$ associated with the Hawing temperature $T$. To probe how the noise and the Hawking radiation collectively affect the uncertainty of interest,
we plot the uncertainty versus Hawking temperature $T$, as shown in Fig. \ref{fig.1}.
These plots show that the magnitude of the uncertainty will increase with growing Hawking temperature in a monotonic way.
This phenomenon is not surprising since the Hawking temperature can reduce the quantum correlation
of $AB$, as depicted in Fig. \ref{fig.1};
To our purpose, we here measure the systemic quantum correlation by means of the so-called quantum discord (QD), which is quantified by
${Q}(\hat{\rho}_{AB})={\cal I}(\hat{\rho}_{AB})-{\cal C}(\hat{\rho}_{AB})$
with mutual information ${\cal I}(\hat{\rho}_{AB})$ and
classical correlation ${\cal C}(\hat{\rho}_{AB})$ \cite{K.C. Tan}. Therefore, this decoherence inevitably results in loss of our ability to guess
the measurement outcome by Alice accurately.

With the current two-particle system considered here, we can derive the QD analytically to be:
\begin{align}
{Q}(\hat{\rho}_{AB})={H_{\rm bin}}({\hat{\rho} _{22}} + {\hat{\rho} _{44}}) + \sum\nolimits_{i = 1}^4 {\chi _i} {\log _2}\chi _i + {\rm min}\{{L}_1, {L}_2\},
\end{align}
with
\[\begin{array}{l}
{L_1} = {H_{\rm bin}}(\xi ),\\
{L_2} = - \sum\nolimits_i {{\hat{\rho} _{ii}}} {\log _2}{\hat{\rho} _{ii}} - {H_{\rm bin}}({\hat{\rho} _{11}} + {\hat{\rho} _{33}}),
\end{array}\]\\
where, $\xi = {{\left\{ {1 + \sqrt {{{\left[ {1 - 2( {\rho _{33}}+{\rho _{44}})} \right]}^2} + 4{{\left( {\left| {{\rho _{14}}} \right| + \left| {{\rho _{23}}} \right|} \right)}^2}} } \right\}} \mathord{\left/
{\vphantom {{\left\{ {1 + \sqrt {{{\left[ {1 - 2({\rho _{44}} + {\rho _{33}})} \right]}^2} + 4{{\left( {\left| {{\rho _{14}}} \right| + \left| {{\rho _{23}}} \right|} \right)}^2}} } \right\}} 2}} \right.
\kern-\nulldelimiterspace} 2}$, ${\hat{\rho} _{ij}}$ is the element of the corresponding density matrix and ${\chi _i}$ denote the eigenstates of ${\hat{\rho} _{AB}}$.

Then, let us consider the lower bound of Eq. (\ref{Eq.3}).
In the bound, $S(A|B_\mathrm{I})=S(\hat{\rho}_{AB_\mathrm{I}})-S(\hat{\rho}_{B_\mathrm{I}})$. While the eigenvalues of $\hat{\rho}_{AB_\mathrm{I}}$ is given by:
\begin{align}
\lambda_{\pm}=(\varepsilon \pm
\tau)/12 , \quad
\eta_{\pm}=(\jmath\pm \ell)/12,
\end{align}
respectively. Where $\varepsilon=3 + a^2 c_3 (3 - 2 p)$,
$\tau=\{(3 - 2 p)^2 + a^4 (3 - 2 p)^2 +
a^2 [9 (-2 + (c_1 - c_2)^2] - 12 (-2 + c_1^2 - 3 c_1 c_2 + 2 c_2^2) p +
4 [-2 + (c_1 - 2 c_2)^2] p^2)\}^{\frac12}$, $\jmath=3 - a^2 c_3 (3 - 2 p)$ and $\ell=\{(3 - 2 p)^2 + a^4 (3 - 2 p)^2 +
a^2 [9 (-2 + (c_1 + c_2)^2] - 12 [-2 + (c_1 + c_2) (c_1 + 2 c_2)] p +
4 [-2 + (c_1 + 2 c_2)^2] p^2\}^{\frac12}$.
Besides, $c$ is preserved to be 1/2 for any pair of Pauli observables. Therefore, one can readily deduce the lower bound is equal to:
\begin{align}
U_b=1-\sum\limits_{i=\pm}\sum\limits_{j=\pm}[\lambda_{i}{\rm log}_2(\lambda_{i})+\eta_{j}{\rm log}_2(\eta_{j})]-H_{\rm bin}(\frac{ a^2 (3 - 2 p) + 2 p}6).
\end{align}

\begin{figure*}
\centering
\includegraphics[width=13cm]{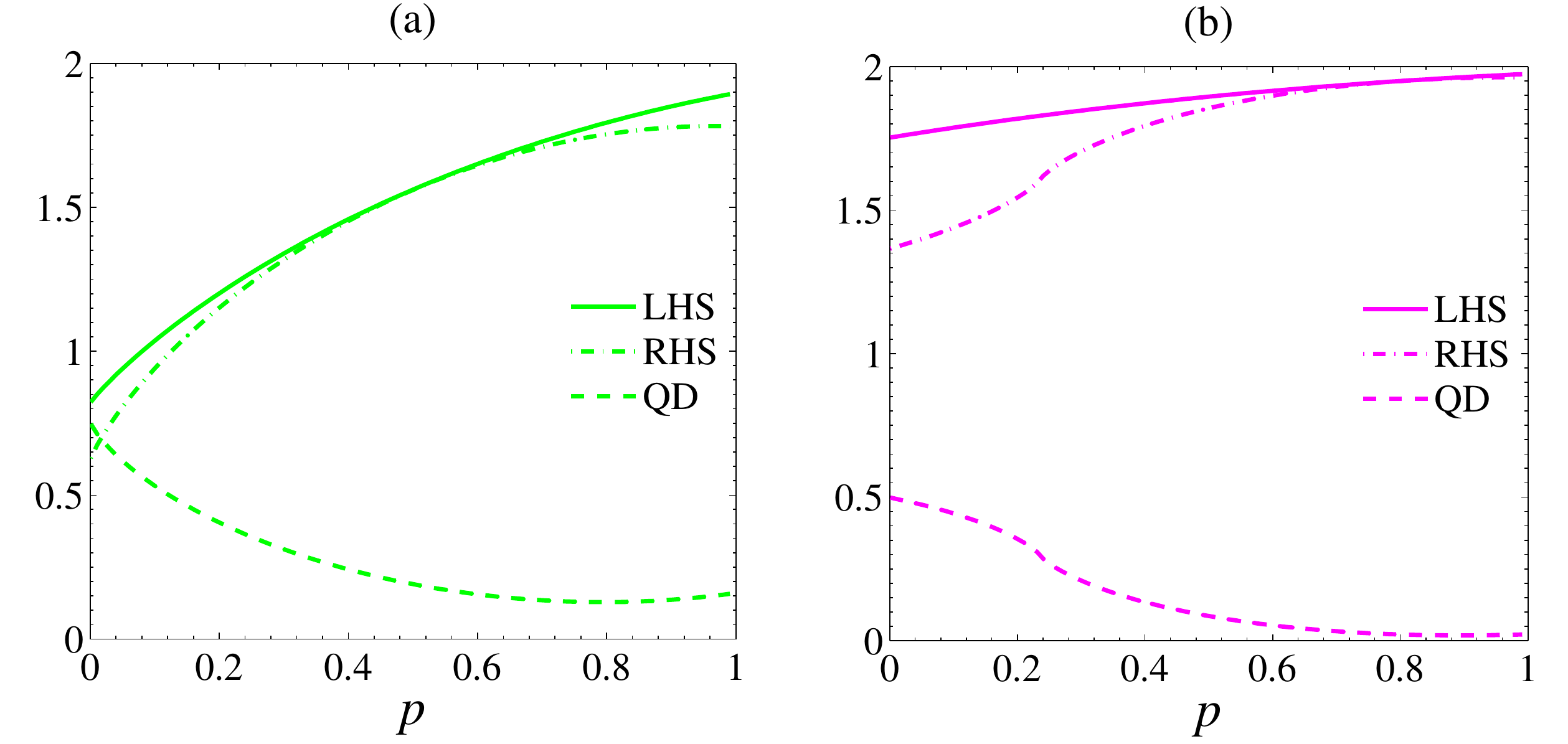}
\caption{Entropic uncertainty and QD with respect to the noise strength, $p$, for different initial states at fixed $T$. All conform to the case where $A$ suffers from DP noise and $B$ (with Hawking temperature $T/\omega=1$) is in the curved space-time. LHS and RHS are in reference to. Eq. (\ref{Eq.3}). QD represents quantum discord of $AB_I$. (a): $(c_1,c_2,c_3)=(1,-1,1)$; Graph (b): $(c_1,c_2,c_3)=(0.5,0.5,0.5)$.}
\label{fig.2}
\end{figure*}

For a maximally entangled state with $(c_1,c_2,c_3)=(1,1,-1)$, ${U= U_b} = 0$ is obtained in the limit of a vanishing Hawking effect and parameterized noise as shown in Fig. \ref{fig.1},
which is in agreement the previous statement. However, as
the Hawking temperature grows larger than zero, the lower bound will increase
and $U> U_b$. For $T\rightarrow\infty$, and the value of the entropic uncertainty will tend towards maximum.

On the other hand, we can also discuss how the noise independently influence the uncertainty. During the evolution of the entropic uncertainty, the lower bound and QD (with respect to the different noise strength, $p$) is plotted for different initial states and fixed Hawking temperature, $T$, given as Figs. \ref{fig.2}(a) and Fig. \ref{fig.2}(b) with initial states $(1,-1,1)$ and $(0.5,0.5,0.5)$, respectively.
From Fig. \ref{fig.2}, we see that the quantum discord will firstly reduce and subsequently recover with increasing noise strength. On the contrary, the uncertainty increases monotonically as the discord increases.
This implies that the measurement uncertainty is not directly synchronous with the systematic quantum correlation (i.e., QD). To further investigate the bound dynamic, we can rewritten the bound into ${U_b}=-{Q}(\hat{\rho}_{AB})+{\rm min}_{\Pi_k^B}[S(\hat{\rho}_{AB}|{\Pi_k^B})]+{\rm log}_2\frac1c $ in terms of
the definition of QD \cite{K.C. Tan}. Based on the above analysis,this behavior
can be interpreted as the uncertainty as it is determined by the QD, $A$'s minimal conditional von Neumann entropy --- ${\rm min}_{\Pi_k^B}[S(\hat{\rho}_{AB}|{\Pi_k^B})]$ --- and dependent on the noise.
In other words, to a large extent the competition of between ${\rm min}_{\Pi_k^B}[S(\hat{\rho}_{AB}|{\Pi_k^B})]$ and ${Q}(\hat{\rho}_{AB})$ determines the dynamics of the uncertainty.

\begin{figure*}
\centering
\includegraphics[width=13cm]{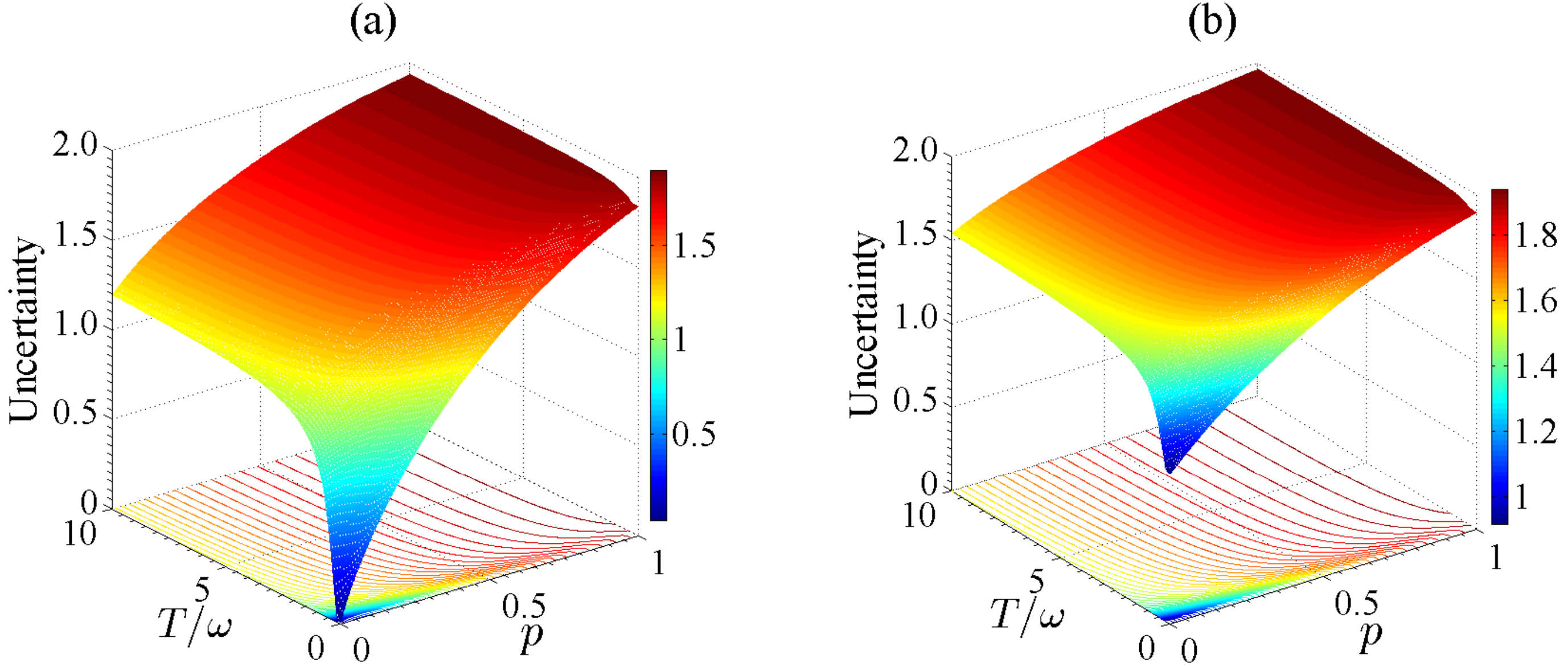}
\caption{The uncertainty as a function of both $p$ and $T/\omega$ in the case of different initial
entanglement prior to being shared between Alice and Bob. Graph (a): the initial state is prepared to (1,-1,1); and
Graph (b): the initial state is limited to (0.7,0.8,0.9).}
\label{fig.3}
\end{figure*}

\begin{figure*}
\centering
\includegraphics[width=13cm]{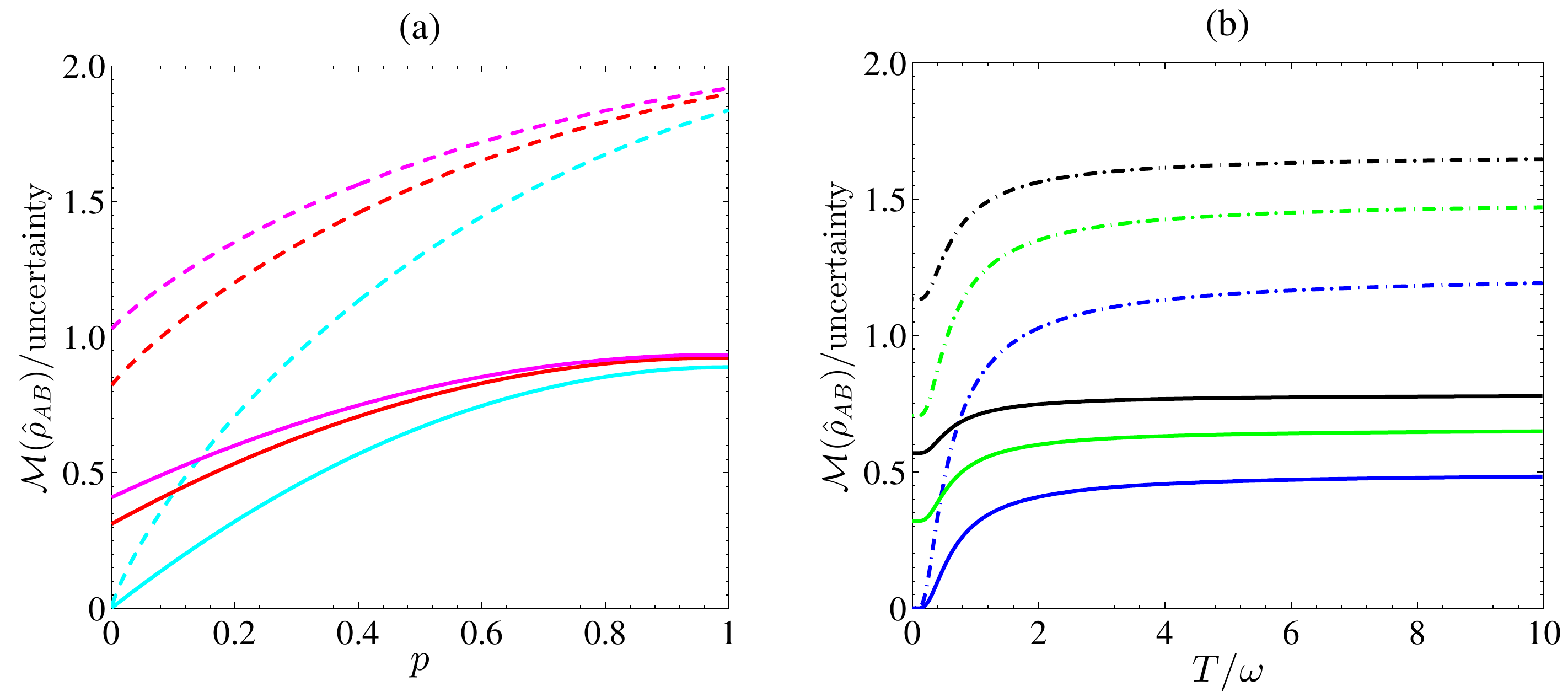}
\caption{The measurement uncertainty and the systematic mixedness ${\cal M}(\hat{\rho}_{AB})$ as a functions of $p$ and $T/\omega$ while the maximally-entangled
initial state $(1,-1,1)$ is prepared. In Graph (a), the cyan, red and magenta lines are cases where $T/\omega=0$, $T/\omega=1$ and $T/\omega=2$, respectively.
The solid lines represents mixedness, ${\cal M}(\hat{\rho}_{AB})$, while the dash lines represents the uncertainty of the measurement;
and in Graph (b), the blue, green and black lines are cases where $p=0$, $p=0.2$ and $p=0.4$.
Similarly, the solid-line denotes mixedness, as the dash-dotted-line denotes the uncertainty of the measurement.}
\label{fig.44}
\end{figure*}

For the sake of disclosing the relationships between the uncertainty, the Hawking temperature ($T$) and the noise strength ($p$), we plot in Fig. \ref{fig.3} these dependencies for initial states
$(1,-1,1)$ and $(0.7,0.8,0.9)$. Fig. \ref{fig.3} straightforwardly indicates that
the greater the Hawking temperature and the greater the noise strength, the larger the uncertainty. If the maximally entangled states are employed as the initial states,
the entropic uncertainty is more dramatically influenced by external noises than by the Hawking radiation, as compared to the mixed initial state. This can be interpreted as the uncertainty being inherently correlated with
the system's mixedness, is denoted as:
\begin{align}
{\cal M}(\hat{\rho}_{AB})=\frac {{d(\hat{\rho}_{AB})}}{{d(\hat{\rho}_{AB})}-1}[1-{\rm Tr}(\hat{\rho}_{AB})^2].
\label{Eq.26}
\end{align}
Within the above, ${d(\hat{\rho}_{AB})}$ is defined as the systematic dimension. To illustrate this conjecture, we plot in Figs. \ref{fig.44}(a) and \ref{fig.44}(b) the measurement uncertainty and the systematic mixedness --- ${\cal M}(\hat{\rho}_{AB})$ --- as functions of the noise strength, $p$, and reduced Hawking temperature, $T/\omega$.
From these figures, one can see that the dynamics of the uncertainty is completely synchronous with the systematic mixedness. In view of the above, it is argued that
the measured uncertainty of interest is strongly correlated with the mixedness ${\cal M}(\hat{\rho}_{AB})$, which is in accordance with the result obtained within Ref. \cite{Z. Y. Xu}.

\subsection{The dynamics of QMA-EUR under collective effects of Hawking radiation field and unital noises}

\begin{figure*}
\centering
\includegraphics[width=13cm]{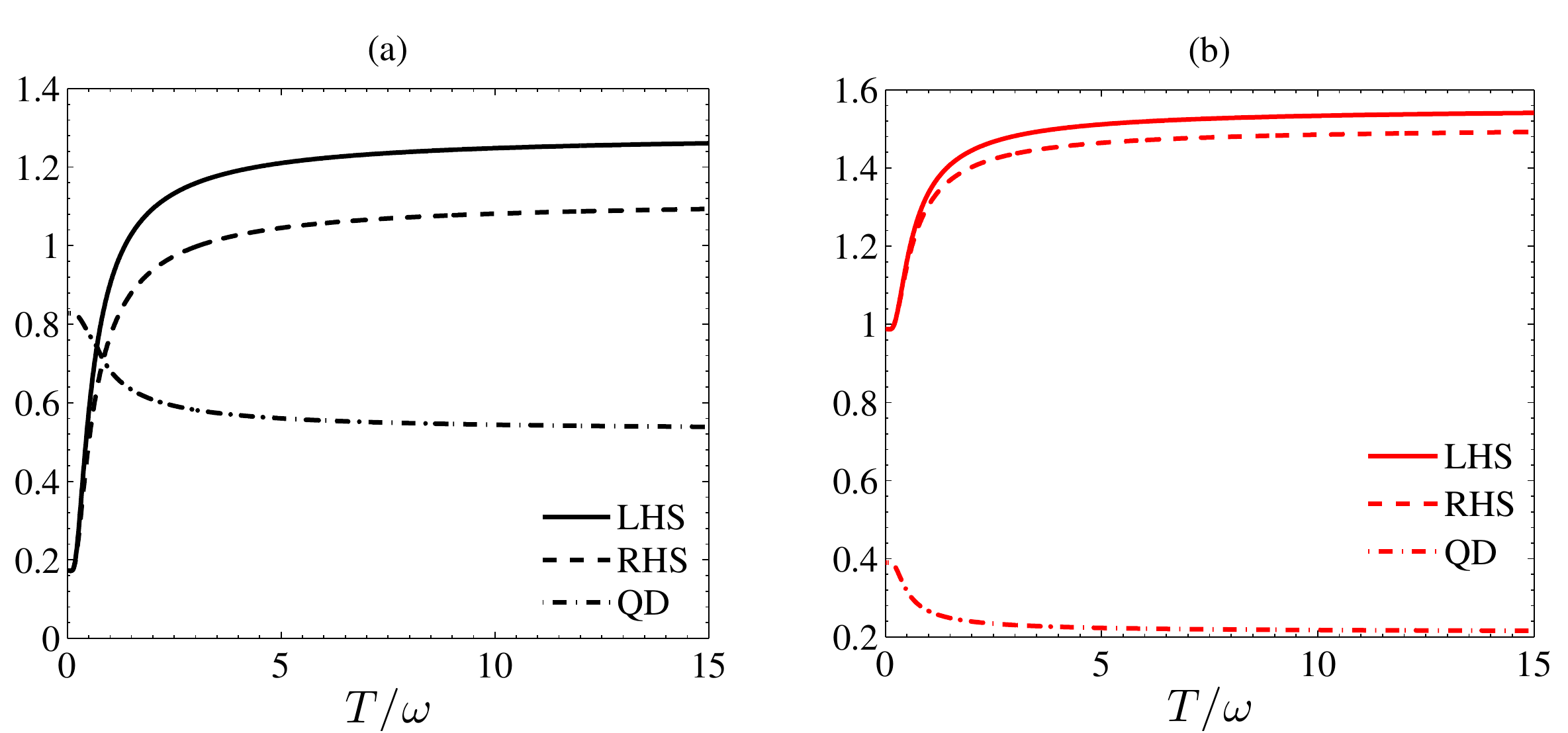}
\caption{A plot of the uncertainty and the QD with respect to $T/\omega$ when $A$ is in a PD
noisy environment with a noise strength of $q=0.1$ and where $B$ is affected by Hawking radiation with a Hawking temperature $T$. LHS and RHS are in reference to the QMA-EUR. Graph (a) is plots with $(c_1,c_2,c_3)=(1,-1,1)$, whereas Graph (b) plots with $(c_1,c_2,c_3)=(0.9,-0.63,0.7)$.}
\label{fig.4}
\end{figure*}

To further investigate this problem, we consider the application of local unital noise on particle $A$. Without lose of generality, we herein take PD noise to be the type of the unital noise, for which the Kraus operators now reads as:
\begin{align}
\hat{\cal K}^0=
\left(
\begin{array}{cc}
1 & 0 \\
0 & \sqrt{1-q}
\end{array}
\right),\quad
\hat{\cal K}^1=
\left(
\begin{array}{cc}
0 & 0\\
0 & {\sqrt q}
\end{array}
\right),
\end{align}
where $q=1-e^{-\Gamma t}$ denotes the decay strength with $q\in[0,1]$ and $\Gamma$ is the decay factor. Therefore,
the dynamics of any particle suffering from PD noise can be mapped into:
\begin{align}
{\hat{\varrho}}_{AB_{\rm I}B_{\rm II}}=\overset{1}{\underset{w=0}\sum} (\hat{\cal K}^w_A\otimes{\mathds{1}}_{B_{\rm I}}\otimes{\mathds{1}}_{B_{\rm II}})\hat{\rho}_{AB_{\rm I}B_{\rm II}}(\hat{\cal K}^w_A\otimes{\mathds{1}}_{B_{\rm I}}\otimes{\mathds{1}}_{B_{\rm II}})^{\dag}.
\label{Eq.27}
\end{align}
Tracing over the degrees in the inaccessible Region II, the systematic state of the bipartite is obtained as:
\begin{align}
{\hat{\varrho}}_{AB_\mathrm{I}}=
\left(
\begin{array}{cccc}
\frac{a^2 (1 + c_3)}4 & 0 & 0 & \frac{ a (c_1 - c_2) \sqrt{(1 - q)} }4\\
0 & \frac{(2 - a^2 (1 + c_3))}4&\frac{ a (c_1 + c_2) \sqrt{(1 - q)} }4& 0 \\
0 & \frac{ a (c_1 + c_2) \sqrt{(1 - q)}}4 & \frac{a^2 (1- c_3)}4 & 0 \\
\frac{ a (c_1 - c_2) \sqrt{(1 - q)} }4& 0 & 0& \frac{ 2 + a^2 (-1 + c_3)}4\\
\end{array}
\right).
\label{Eq.29}
\end{align}
Because Alice is free from Hawking effect, the measurement result
should be independent of particle $B$'s state.
Nevertheless, it notes that the conditional von Neumann entropies will be
transformed; the below formulae offer new post-measured states:
\begin{align}
\hat{\varrho}_{\hat{\sigma}_xB_\mathrm{I}}=&a^2/4(|00\rangle\langle00|+|10\rangle\langle10|)+(1/2 - a^2/4)(|01\rangle\langle01|+|11\rangle\langle11|)\nonumber \\
&+1/4 a c_1 \sqrt{1 - q}(|01\rangle\langle10|+|10\rangle\langle01|+|00\rangle\langle11|+|11\rangle\langle00|),\nonumber \\
\hat{\varrho}_{\hat{\sigma}_zB_\mathrm{I}}=&1/4 a^2 (1 + c_3)|00\rangle\langle00|+1/4 (2 - a^2 (1 + c_3))|01\rangle\langle01| \nonumber \\
&-(1/4) a^2 (-1 + c_3)|10\rangle\langle10|+1/4 (2 + a^2 (-1 + c_3))|11\rangle\langle11|.
\label{Eq.22}
\end{align}
After obtaining the eigenvalues of $\hat{\varrho}_{\hat{\sigma}_xB_\mathrm{I}}$ and $\hat{\varrho}_{\hat{\sigma}_zB_\mathrm{I}}$,
the entropies of the post-measured states are calculated as:
\begin{align}
H(\hat{\varrho}_{\hat{\sigma}_xB_\mathrm{I}})=H_{\rm bin}(\frac{1+\kappa}2)+1, \quad
H(\hat{\varrho}_{\hat{\sigma}_zB_\mathrm{I}})=-\sum_{i=1}^4\kappa_{i} {\rm log}_2\kappa_{i},
\end{align}
with $\kappa=\sqrt{1 + a^4 -a^2 [2 +{c_1}^2 (q-1)]}$,
$\kappa_1=a^2 (1 + c_3)/4$, $\kappa_2= [2 - a^2 (1 + c_3)]/4$, $\kappa_3=a^2 (1 - c_3)/4$ and
$\kappa_4=[2 + a^2 (-1 + c_3)]/4$.

\begin{figure*}
\centering
\includegraphics[width=13cm]{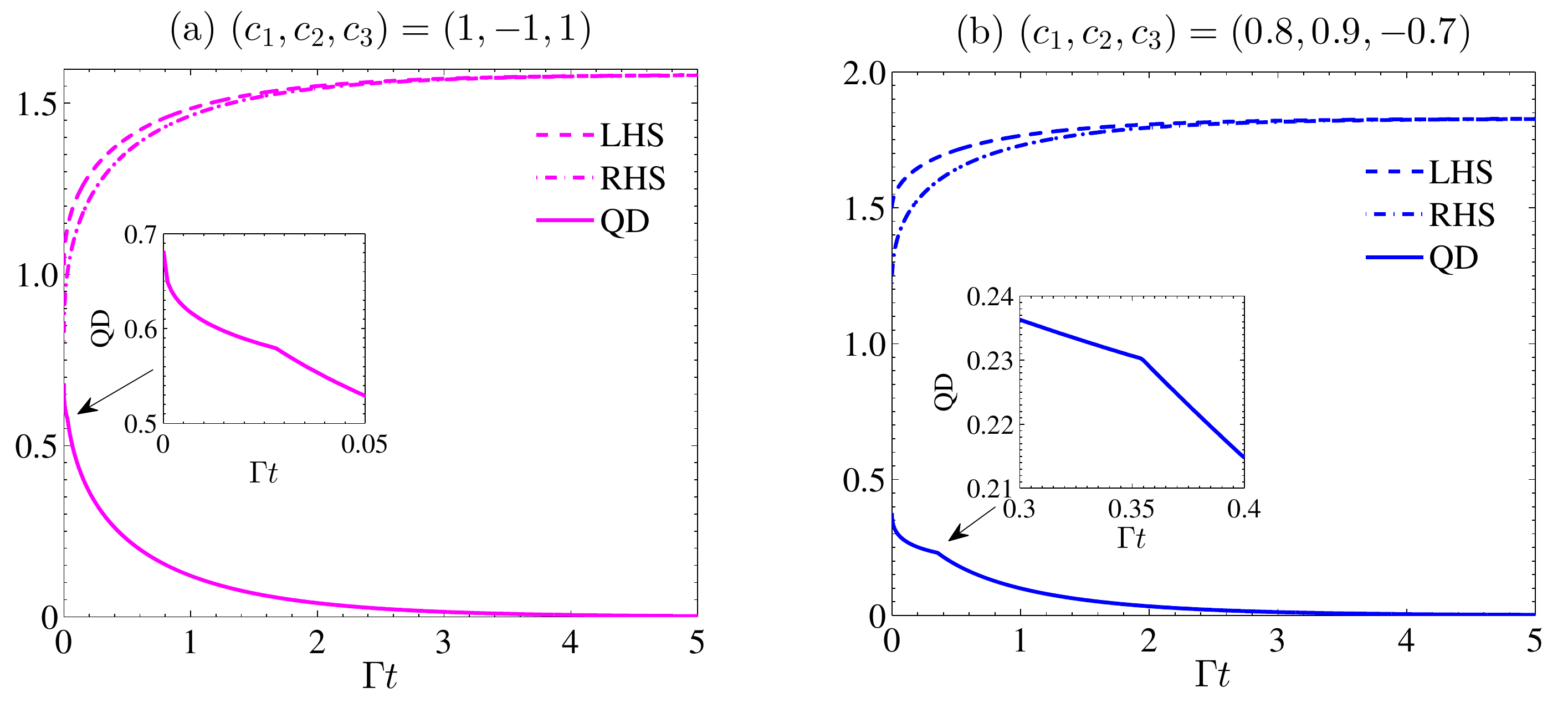}
\caption{The QD and the uncertainty as a functions of $\Gamma t$ in different initial states while $A$ experiences DP noises and $B$ experiences Hawking temperature $T/\omega=2$. Graph (a):
$(c_1,c_2,c_3)=(1,-1,1)$, and Graph (b): $(c_1,c_2,c_3)=(0.8,0.9,-0.7)$.}
\label{fig.5}
\end{figure*}

\begin{figure*}
\centering
\includegraphics[width=7cm]{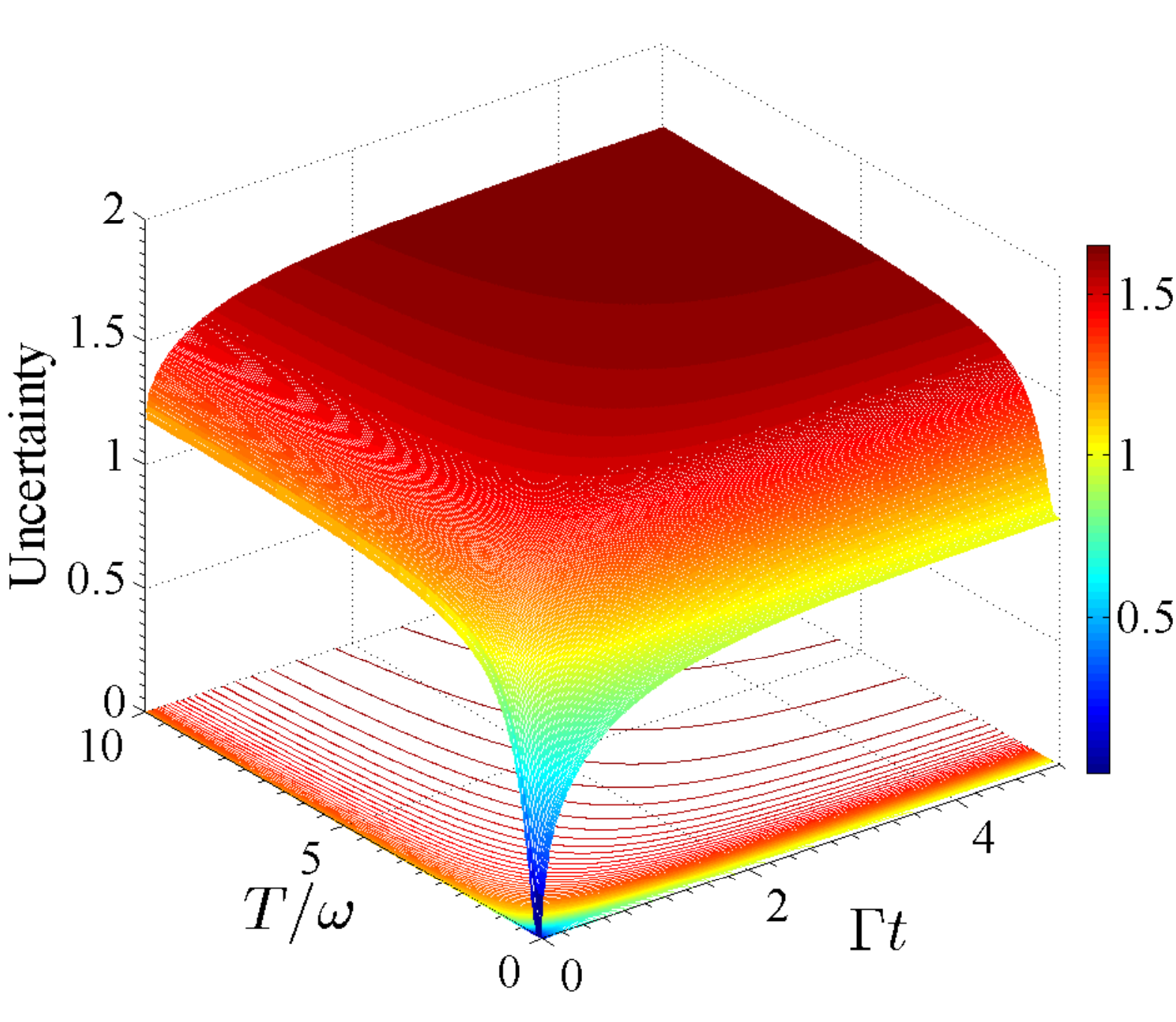}
\caption{QMA-EUR versus both $\Gamma{t}$ and $T/\omega$ with regard to the initial states' parameter $(c_1,c_2,c_3)=(1,-1,1)$.}
\label{fig.7}
\end{figure*}

\begin{figure*}
\centering
\includegraphics[width=13cm]{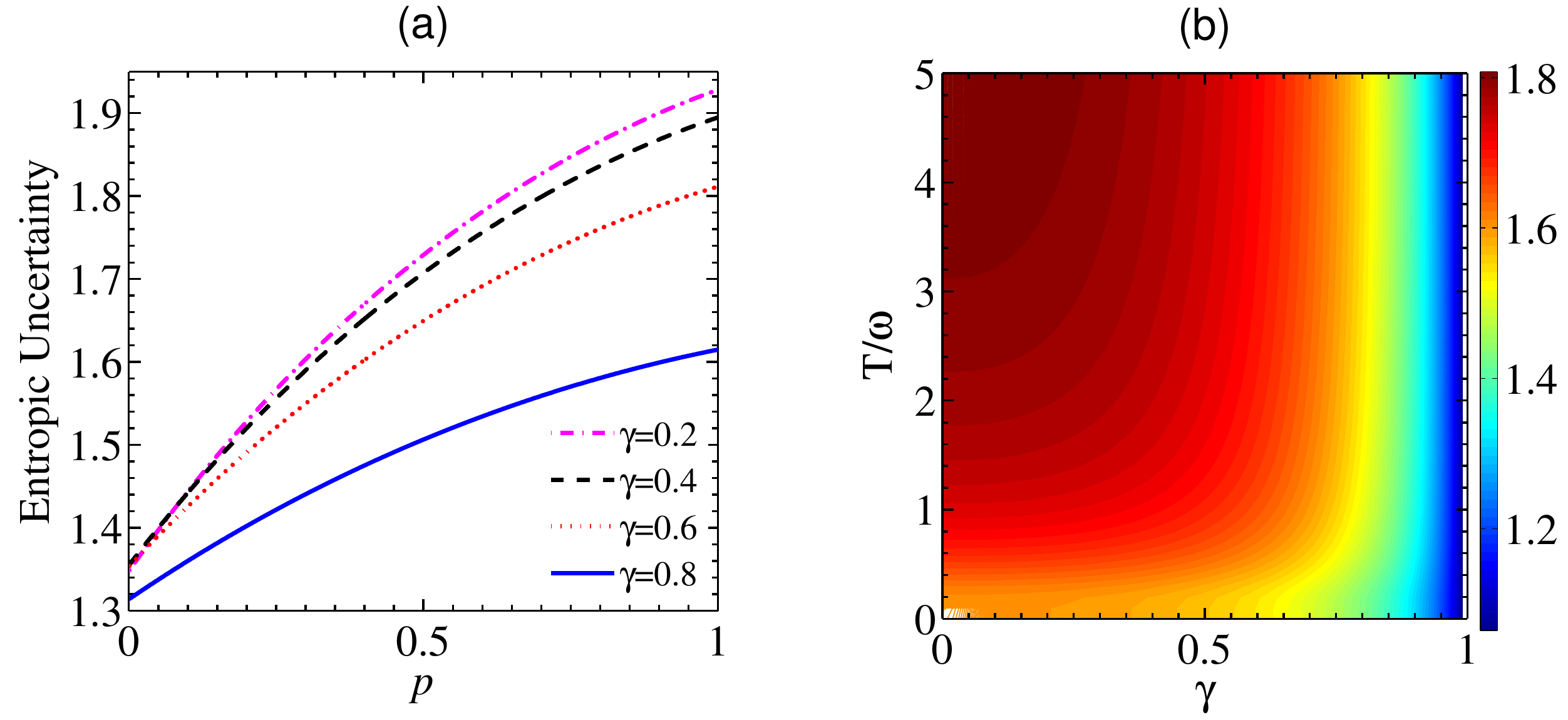}
\caption{The uncertainty as a function of the weak measurement strength $\gamma$, the Hawking temperature $T/\omega$ and the decoherence strength $p$ in the case of DP noise. In Graph (a), the operation strength $\gamma$ changes from 0.2 to 0.8 from top to bottom and $T/\omega=1$. In graph (b), the measurement uncertainty varies with the weak measurement strength $\gamma$ and $T/\omega$ with $p=0.5$.
For all plotted with the initial states' parameter $(c_1,c_2,c_3)=(0.9,-0.8,0.6)$.}
\label{fig.8}
\end{figure*}

\begin{figure*}
\centering
\includegraphics[width=13cm]{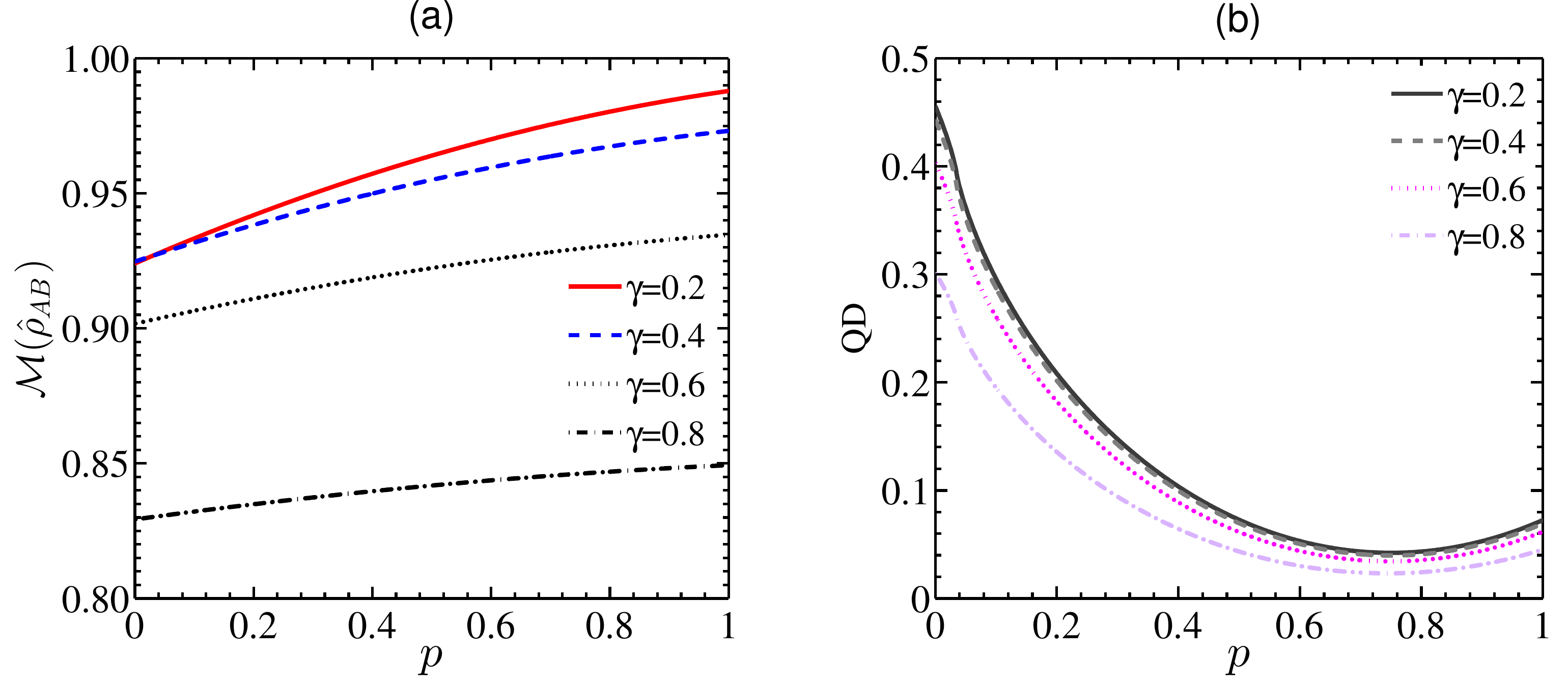}
\caption{The systematic mixedness (${\cal M}(\hat{\rho}_{AB})$) and QD as a function of the decoherence strength, $p$, for different weak measurement strengths in the case of DP noise. In Graphs (a) and (b), the operation strength $\gamma$ changes from 0.2 to 0.8 from top to bottom, $T/\omega=1$ and the states' parameters satisfy $(c_1,c_2,c_3)=(0.9,-0.8,0.6)$.}
\label{fig.88}
\end{figure*}

Because $H(\hat{\varrho}_{B_\mathrm{I}})=H_{\rm bin}(\frac{a^2}2)$, we may derive that the measured uncertainty in such a scenario is quantified by:
\begin{align}
\widetilde{U}=H_{\rm bin}(\frac{1+\kappa}2)-\sum\limits_{i=1}^4\kappa_{i} {\rm log}_2\kappa_{i}-2H_{\rm bin}(\frac{a^2}2)+1 ,
\end{align}
which evolution is depicted in Fig. \ref{fig.4}. It is easy to see that the uncertainty increases with the Hawking temperature, $T$, when the decay strength is constrained to $q=0.1$,
and then saturates into a peak value with large enough $T$. This shows that Particle $B$ --- located in the thermal field --- will not act on the measurement uncertainty
if $T$ is sufficiently large;
this is because the quantum correlation between $A$ and $B_{\rm I}$
will be frozen, leading to an extreme asymptotically frustration of information flow
among the two correlated particles, the external noise environment and the physical inaccessible Region II in the radiation field, as verified by Ref. \cite{S. Xu}.

With regard to the uncertainty bound, the first term is $S(A|B_\mathrm{I})=S(\hat{\varrho}_{AB_\mathrm{I}})-S(\hat{\varrho}_{{B_{\mathrm{I}}}})$.
The eigenvalues of $\hat{\varrho}_{AB_{\mathrm{I}}}$ are:
\begin{align}
\iota_{\pm}=(x\pm y)/16, \quad
\zeta_{\pm}=(\varsigma\pm \upsilon)/16,
\end{align}
where $x=1 - a^2 c_3$, $y=\{1 + a^4 + a^2 [-2 - (c_1 + c_2)^2 (-1 + q)]\}^{\frac12}$, $\varsigma=1 + a^2 c_3$ and $\upsilon=\{1 + a^4 + a^2 [-2 - (c_1 - c_2)^2 (-1 + q)]\}^{\frac12}$. After some calculations,
we ascertain that the bound shall take the analytical form:
\begin{align}
\widetilde{U}_b=-\sum\limits_{i=\pm}\sum\limits_{j=\pm}(\iota_{i}{\rm log}_2\iota_{i}+\zeta_{j}{\rm log}_2\zeta_{j})-H_{\rm bin}(\frac{a^2}2)+1.
\end{align}
As depicted in Fig. \ref{fig.4} for a maximally-entangled initial state characterized by $(c_1,c_2,c_3)=(1,-1,1)$, for $q\neq0,1$, the bipartite $AB$ will be in a non-maximal entanglement for all values of the Hawking temperature,
$U>U_b$; thereby satisfying the QMA-EUR, shown in Eq. (\ref{Eq.3}). Notably, as
$T\neq0$ is held, the uncertainty bound will raise. Similarly,
the measurement uncertainty would increase to a fixed peak value if the Hawking temperature is large enough.

Additionally, the evolutions of the QD and the uncertainty are both provided in Fig. \ref{fig.5} with growing the decay time $\Gamma t$ when $T/\omega$ is taken to be a fixed value.
In Fig. \ref{fig.5}, one can understand that the quantum correlation does not always undergo smooth evolution with an increasing decay time, and that there exists a singlet during the
evolution. By contrast, the uncertainty grows monotonically with $\Gamma t$, which further proves our previous statement that the measurement uncertainty is not entirely
synchronous with systematic quantum discord. At the same time, we draw the evolution of the uncertainty versus
$\Gamma t$, and the temperature, $T/\omega$, in Fig. \ref{fig.7}.
This figure shows that the uncertainty can be increased to a finite maxima with both a growing decay time and a greater Hawking temperature.
As previously stated, the mixedness ${\cal M}({\hat \varrho}_{AB})$ can be easily obtained by combining Eqs. (\ref{Eq.26}) and (\ref{Eq.29}).
The same result can be inferred as the measured uncertainty is strongly relative to the systemic mixedness.

Actually, if the situation were reversed, the measured particle located near the event horizon while the memory interacts with its surrounding environment. In this situation, one can say that firstly, the Hawking radiation will destroy the entanglement between A and B, which will inevitably result in information outflow from the system into the inaccessible Region II in the Schwarzschild space-time, and the inflation of the measurement uncertainty as well. Secondly, the noise affecting B --- including the unital and nonunital noise --- will also lead to the decay of the system; thus, the systematic quantum correlation will decrease. With this knowledge in hand, we conclude that the information will outflow into the environment and the uncertainty will raise naturally. Thirdly, we infer that the Hawking effect might become a prominent factor affecting the dynamics of the measurement uncertainty when compared with the environmental noises; this is in stark contrast to the scenario in the current consideration.

\section{Steering the uncertainty via an uncollapsed measurement}

\begin{figure*}
\centering
\includegraphics[width=13cm]{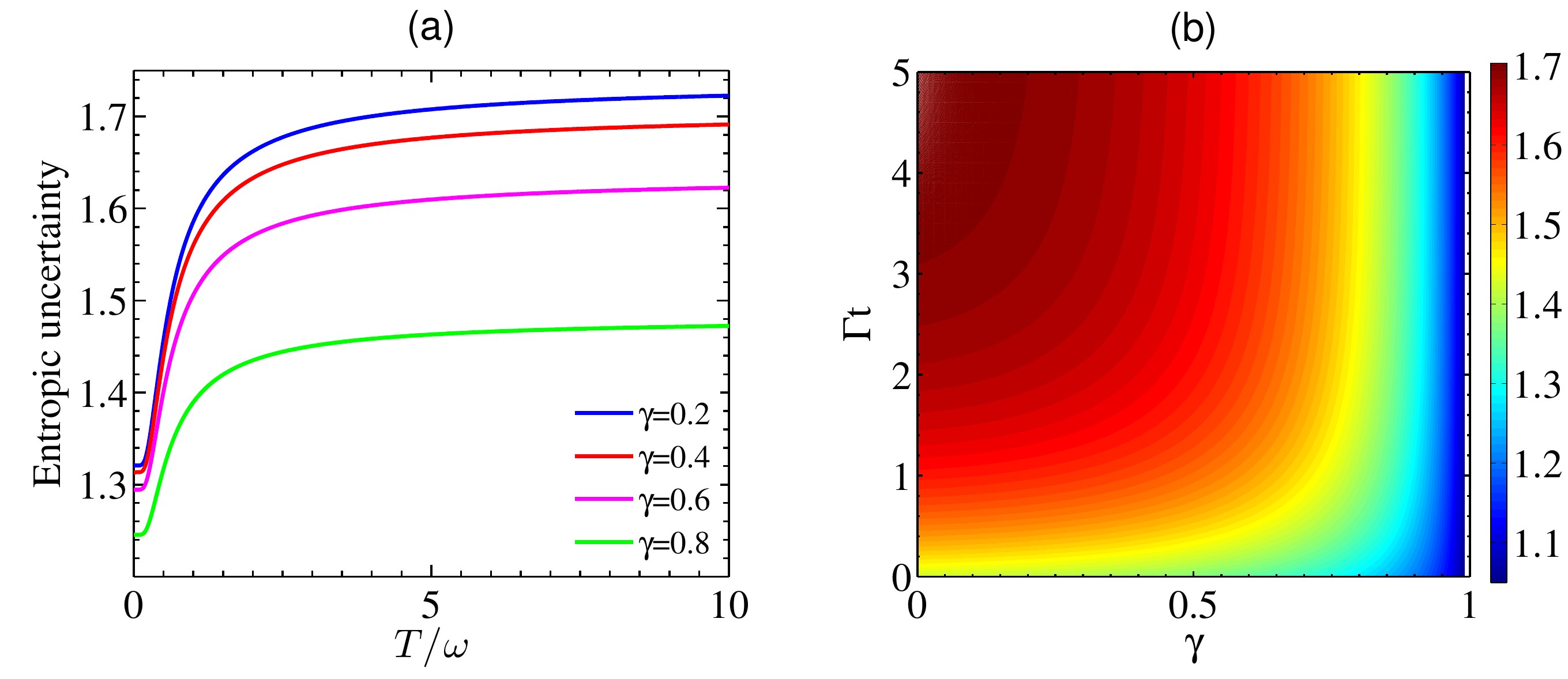}
\caption{The QMA-EUR as a function of weak measurement strength ($\gamma$), the decoherence time ($\Gamma t$) and Hawking temperature ($T/\omega$) in the case of PD noise. In Graph (a), the operation strength ($\gamma$) ranges from 0.2 to 0.8, from top to bottom; the decoherence strength is set to $q=0.6$. Graph (b), the measurement uncertainty varies with the weak measurement strength ($\gamma$) and the decoherence time ($\Gamma t$) with a fixed temperature $T/\omega=1$.
All plots used initial state parameter $(c_1,c_2,c_3)=(0.7,0.6,-0.8)$.}
\label{fig.9}
\end{figure*}

In the context of practical information-processing, it is in demand to achieve a relatively small degree of measurement uncertainty; therefore, an optimally small measurement uncertainty is crucial for achieving practical quantum tasks related to quantum measurements to reduce the uncertainty of interest.
Motivated by this, we herein put forward a simple functional strategy
to control the magnitude of the entropic uncertainty and to suppress decoherence via employing a class of uncollapsing operations, \emph{i.e.} quantum weak measurement (QWM) \cite{S. C. Wang,X. Xiao,Y. Aharonov}.
Mathematically, a QWM process can be mapped into a matrix with form of:
\begin{align}
{\hat{ M}}^{\rm wm}=\left(
\begin{array}{cc}
1 & 0 \\
0 & \sqrt{1-\gamma}
\end{array}
\right),
\end{align}
where $\gamma$ represents the measurement strength, $\gamma \in[0,\ 1]$. In fact, QWM is designed to accomplish a postselection that
performs the transition $|1\rangle\rightarrow|0\rangle$; the operation can be achieved based on an ideal detector
to supervise laboratory environments. Specifically, it is conventionally regarded as null-result QWM on account that the detector
never gives a signal. With respect to QWM, thorough collapse into the corresponding eigenstates will not take place, hence the particles continue their own evolution.
The merit of quantum weak measurements is in an effective suppression of quantum decoherence by uncollapsing the systematic state and
prompting the system to an excited state. Through performing the QWM on particle $A$, the state explicitly becomes:
\begin{align}
{\tilde{\rho}}_{AB_{\rm I}}^{\rm wm}(t)={({\hat{ M}}^{\rm wm}_A\otimes {{\mathds{1}}}_{B_{\rm I}})\hat{\rho}_{A{B_{\rm I}}}{({\hat{ M}}^{\rm wm}_A\otimes {\mathds{1}}_{B_{\rm I}})}^{\dag}}/{\cal P}_{succ}.
\end{align}
where ${\cal P}_{succ}={{\rm Tr}_{AB_{\rm I}}[{({\hat{M}}^{\rm wm}_A\otimes {\mathds{1}}_{B_{\rm I}})\hat{\rho}_{A{B_{\rm I}}}{({\hat{ M}}^{\rm wm}_A\otimes {\mathds{1}}_{B_{\rm I}})}^{\dag}}]}$
quantifies the success probability
of the measurement.

First of all, let us examine the effect of the QWM on QMA-EUR under the Hawking radiation and DP noises.
After calculations, we can obtain the post-measurement state with the matrix elements
\begin{align}
\tilde{\rho}_{11}=&{ N}(-2 p + a^2 (1 + c_3) (-3 + 2 p)),\nonumber \\
\tilde{\rho}_{22}=&-{N}(6 - 2 p + a^2 (1 + c_3) (-3 + 2 p)),\nonumber\\
\tilde{\rho}_{33}=&{N}(( \gamma-1) (2 p + a^2 (-1 + c3) (-3 + 2 p))),\nonumber\\
\tilde{\rho}_{44}=&{ N}((1-\gamma) (2 (-3 + p) + a^2 (-1 + c_3) (-3 + 2 p)),\nonumber\\
\tilde{\rho}_{14}=&{\tilde{\rho}}_{41}={ N}(a \sqrt{1 - \gamma} (c_2 (3 - 4 p) + c_1 (-3 + 2 p))),\nonumber\\
\tilde{\rho}_{23}=&{\tilde{\rho}}_{32}={ N}(a \sqrt{1 - \gamma} (-3 (c_1 + c_2) + 2 (c_1 + 2 c_2) p)),
\end{align}
where other elements are zero-valued; the normalized coefficient is ${N}=1/{(6 ( \gamma-2))}$. And thus we obtain that the entropies can be given by:
\begin{align}
\tilde{H}({\hat{\rho}}_{\hat{\sigma}_xB_\mathrm{I}})=H_{\rm bin}(\frac{1+4\tilde{\lambda}}2)+1, \quad
\tilde{H}({\hat{\rho}}_{\hat{\sigma}_zB_\mathrm{I}})=-\sum_{i}\tilde{\rho}_{ii} {\rm log}_2\tilde{\rho}_{ii},
\end{align}
respectively. Where $\tilde{\lambda}=\{((-2 +\gamma)^2 + a^4 (2 + (-1 + c_3) \gamma)^2 -
2 a^2 (2 c_1^2 (-1 + \gamma) - (-2 + \gamma) (2 + (-1 + c_3) \gamma))) (3 -
2 p)^2\}^{\frac 12}/(12 (-2 + \gamma))$. And the von Neumman entropy of the reduced matrix for $B_{\rm I}$ is taken as:
$\tilde{H}({\hat{\rho}}_{B_\mathrm{I}})=H_{\rm bin}(\Delta)$
with $\Delta=\frac{2 (-2 + \gamma) p + a^2 (2 + (-1 + c_3) \gamma) (-3 + 2 p)}{6 (\gamma-2 )}$. As a consequence, the uncertainty can be
quantified by:
\begin{align}
U^{\rm wm}_{DP}=H_{\rm bin}(\frac{1+4\tilde{\lambda}}2)-\sum_{i=1}^4\tilde{\rho}_{ii} {\rm log}_2\tilde{\rho}_{ii}+2H_{\rm bin}(\Delta)+1.
\end{align}

For the sake of clarity, we show the relationship between the entropic uncertainty and the measurement strength $\gamma$ in
Fig. \ref{fig.8}(a); the initial state's parameters $(c_1,c_2,c_3)=(0.9,-0.8,0.6)$ were employed. From Fig. \ref{fig.8}(a), we can
readily infer the measurement uncertainty will be reduced as $\gamma$ grows. Moreover, we draw --- in Fig. \ref{fig.8}(b) --- the relationship among the Hawking temperature, the measurement strength and the entropic uncertainty; the uncertainty increases with growing temperature and decreases with an increase in
measurement strength. This is essentially in agreement with the results we previously stated.

In this case, the systematic mixedness can be calculated as:
\begin{align}
{\cal M}(\hat{\rho}_{AB})=&\frac{4}{3}\{1-[(\gamma-1)^2 (a^2 ({c_3}-1) (2 p-3)+2 (p-3))^2+(\gamma-1)^2 (a^2 ({c_3}-1) (2 p-3)+2 p)^2 \nonumber\\
&+(a^2 ({c_3}+1) (2 p-3)-2 p)^2+(a^2 ({c_3}+1) (2 p-3)-2 p+6)^2)/(36 (\gamma-2)^2]\}.
\end{align}
Additionally, we plot the mixedness and quantum discord as functions of the decoherence strength, $p$, for different weak measurement parameters, $\gamma$, as shown in Fig. \ref{fig.88}. It shows that
the mixedness is highly synchronous with the uncertainty, as indicated in Figs. \ref{fig.8}(a) and \ref{fig.88}(a), and the
stronger the weak measurement, the smaller mixedness of the composite system of interest.
By contrast, the stronger weak measurement will bring on the smaller quantum correlation of the system as shown in \ref{fig.88}(b).

Next, let us consider the effect of the QWM on QMA-EUR under unital noise (PD). Likewise,we derive a form for the entropic uncertainty:
\begin{align}
{U}^{\rm wm}_{PD}=H_{\rm bin}(\frac{1+\tilde{\kappa}}2)-\sum\limits_{i=1}^4\tilde{\kappa}_{i} {\rm log}_2\tilde{\kappa}_{i}-2H_{\rm bin}(\frac{a^2 (\gamma-2 - c_3 \gamma)}{2 (\gamma-2 )})+1,
\end{align}
where $\tilde{\kappa}=\{{(\gamma-2)^2 + a^4 (2 + (c_3-1) \gamma)^2 + 2 a^2 (2 c_1^2 ( q-1) (\gamma-1) + (\gamma-2) (2 + ( c_3-1) \gamma))}\}^{\frac12}/(\gamma-2)$, $\tilde{\kappa}_{1}=\frac{a^2 (1 + c_3)}{2 (2 - \gamma)}$,
$\tilde{\kappa}_{2}=\frac{2 - a^2 (1 + c_3)}{2 (2 - \gamma)} $, $\tilde{\kappa}_{3}= \frac{a^2 (c_3-1 )( \gamma-1)}{2 (2 - \gamma)}$ and $\tilde{\kappa}_{4}= \frac{((2 + a^2 (-1 + c_3)) (1- \gamma)}{2 (2 - \gamma)}$.
We provide Fig. \ref{fig.9}(a) to show the uncertainty between the reduced Hawking temperature, $T/\omega$, and the entropic uncertainty with different measurement strength
with the PD decoherence strength $q=0.6$ and initial state parameters $(c_1,c_2,c_3)=(0.7,0.6,-0.8)$. From Fig. \ref{fig.9}(a), it has been shown that the uncertainty will decrease with an increase in measurement strength, $\gamma$. We draw the relationship among the decoherence time,
the measurement strength and the entropic uncertainty in Fig. \ref{fig.9}(b); the uncertainty increases with the decoherence time and reduces with a growing
measurement strength. In other words, the weak measurement is working on the reduction of the measured uncertainty, highly requested in
the regime of realistic quantum information processing.

\section{Summary and conclusions}

\begin{figure*}
\centering
\includegraphics[width=8cm]{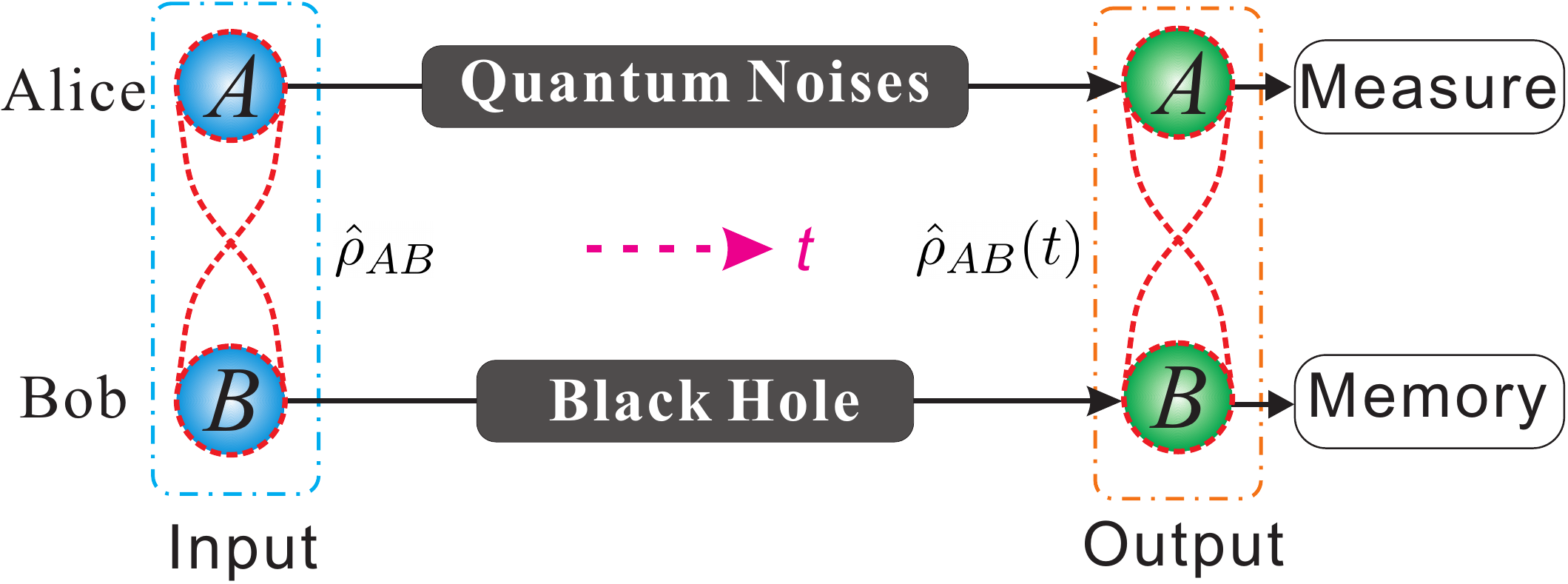}
\caption{A schematic diagram for the probed model with $A$ in flat time-space and $B$ --- taking on a quantum memory reservoir --- in the Schwarzschild space-time.
$A$ remains in an inertial frame, while
$B$ exists near the event horizon of Schwarzschild black hole.}
\label{fig.10}
\end{figure*}

In this study, we have examined the dynamical characteristic of the uncertainty via entropy within the real-world architecture, as diagrammatized in Fig. \ref{fig.10}:
Particle $A$ (at Alice's location) is subjected to a noisy environment and $B$ --- serving as a quantum memory reservoir --- stays near the event horizon
in a Schwarzschild space-time, exposed to Hawking radiation.
Explicitly, we study the dynamical evolutions of the uncertainty in the course of quantum measurements consisting of incompatible observables
over collective effects of Hawking effect and exposed to either nonunital (DP) or unital (PD) noise. We verify that the
Hawking effect --- from the thermal radiation field interacting with the quantum memory particle --- gives rise to reduction of quantum correlation for the bipartite $A-B$.
Thereby, an increase in the measurement uncertainty is observed within Region $\mathrm{I}$;
this physical phenomenon can be explained by a information redistribution of the whole system, in which some valid information flows into the physical inaccessible Region {\rm II}.
Noteworthy, in the high-temperature region, the measurement uncertainty inflates toward a fixed asymptotic value.
Moreover, it was determined that in the presence of depolarizing noise, the QD firstly reduces non-monotonically as the noise strength ($p$) grows, then recover to a degree.
Interestingly, the uncertainty of interest increases monotonically with increasing $p$. This reflects the fact that the QD is not the determining factor, and we deduce
the amount of the uncertainty is also dependent of the conditional von Neumman entropy for subsystem to be measured.
For a unital (PD) noise, the entropic uncertainty monotonically increases with the increasing decay time, $\Gamma t$; that implies,
the information outflows from the system of interest to the noisy environment, and not return.
Contrarily, the quantum correlation will decrease with time at finite Hawking temperatures, in spite of the existence of a singlet during the evolution.
Furthermore, we argue that the measured uncertainty is strongly associated with the systematic mixedness;
it turns out that a smaller mixedness can lead to the less measurement uncertainty.
In final, we design a methodology to reduce the magnitude of the measurement uncertainty by quantum weak measurement, considerably requested in measurement-based information processing.

Additionally, we note that there exist two previous literatures related to this field. Although the previous and the current works all observe the dynamics of
the measured uncertainty in the Schwarzschild black hole, there exists some apparent differences: firstly, the scenarios considered are very different, Refs. \cite{Jun Feng2} and \cite{huang} investigated the scenario that the measured particle, $A$, is located near the event horizon while the memory, $B$, is free from any environment. By contrast, our investigations mainly concentrate on a distinct scenario where the particle to be measured stays at a flat space-time and the memory hovers near the event horizon, the inverse scenario. Secondly, the influence factors of the uncertainty under consideration are very distinct. Refs. \cite{Jun Feng2} and \cite{huang} examined the relationship between the uncertainty and the distance between Bob and the event horizon, the mode frequency of quantum memory and the mass of black hole, and \cite{huang} demonstrated a tighter bound via the Holevo quantity. While we contribute to unveiling how the Hawking radiation, and the noisy strength of various types of noise influence the uncertainty. Furthermore, we design a working strategy to reduce the measurement uncertainty, while this is not considered in \cite{Jun Feng2} and \cite{huang}.

To sum up, we believe that our observations might be helpful to better understand the dynamic characters of the measurement uncertainty
under a curved space-time, and also be nontrivial for quantum measurements during relativistic quantum information sciences.

\begin{acknowledgments}
This work was supported by NSFC (Grant Nos. 61601002 and 11575001), the Fund from CAS Key Laboratory of
Quantum Information (Grant No. KQI201701), and Anhui Provincial Natural Science Foundation (Grant No. 1508085QF139).
\end{acknowledgments}

\end{document}